\documentstyle[12pt]{article}

\input epsf

\def\kms{km~s$^{-1}$}
\newcommand{\lsim}{\ \raise -2.truept\hbox{\rlap{\hbox{$\sim$}}\raise5.truept
	\hbox{$<$}\ }}
\newcommand{\gsim}{\ \raise -2.truept\hbox{\rlap{\hbox{$\sim$}}\raise5.truept
	\hbox{$>$}\ }}

\begin{document}


\begin{center}
{\bf  COMPARISON OF STARS AND DECAYING NEUTRINOS }

{\bf AS ADDITIONAL SOURCES OF INTERGALACTIC UV BACKGROUND}

\vspace{.4cm}
{\small \bf Luis Masperi}\\

{\it Centro At\'omico Bariloche and Instituto Balseiro,
 Comisi\'on Nacional de Energ\'{\i}a At\'omica and Universidad Nacional
de Cuyo,
 8400  S.C. de Bariloche, Argentina}\\

\vspace{.2cm}
and
\vspace{.2cm}

{ \small \bf Sandra Savaglio}\\
{\it European Southern Observatory, Karl-Schwarzschildstr. 2,
	Garching bei M\"unchen, D--85748 Germany}
\end{center}


\begin{abstract}
A numerical calculation of properties of finite absorbers and of intergalactic
medium based on photoionization equilibrium is performed to confront
alternative UV sources in addition to quasars.

It is seen that a spectrum including a large peak around the HI ionization
energy due to decaying neutrinos is too soft in the region up to the HeI
edge to explain the relatively small observed ratio of neutral He and H
densities in Lyman-limit systems if their size is of the kpc order. The
recently proposed decrease of the contribution from unstable neutrinos
solves this problem but tends to spoil the consistence between IGM and
Lyman-$\alpha$ clouds, which requires a large ratio of fluxes for the HI and HeII
ionization frequencies, unless there is a very fast decline of quasars above
$z = 3$.

On the other hand, the addition of stars to quasars may produce a spectrum
sufficiently hard between HI and HeI and thereafter soft up to HeII to
allow a reasonable agreement of the properties of denser absorbers with those
of IGM. This model seems to favour cold dark matter with additional
cosmological constant.

\end{abstract}


 \vspace{0.6cm}

\noindent {\bf 1. Introduction}

\vspace{0.2cm}
The reionization of the Universe
is a subject of deep
experimental and theoretical study for its implication on  galaxy formation
 and on the consistence of small scale anisotropies with
cosmological models. The different  possible sources of UV background may
give some information on the identity of dark matter.

It has been seen from the bounds on Gunn-Peterson (GP) effect (Gunn \&
Peterson 1965) for neutral
H and single-ionized He that the frequency UV spectrum
seems to be softer than that due to quasars (QSO) (Madau 1992), requiring
therefore additional sources which might be stars.

Another suggested possibility has been that
decaying neutrinos (Sciama 1990a), in addition to QSOs, may ionize HI in
the intergalactic medium (IGM), clouds and other systems. This decaying dark
matter (DDM) would be hot (HDM) and might be in difficulty to explain the
formation of structures unless cosmic strings are the seed of primeval
fluctuations (Zanchin et al. 1996).
 Since this is a delicate subject, it is interesting to see whether
 the sole
properties of reionization are able to support  this hypothesis or not.

The observations whose validity we will assume for our analysis are the
ratio of densities of neutral He and H in
Lyman-limit systems (LLS) (Reimers \& Vogel 1993), the bound of GP effect
for HI (Giallongo et al. 1994) considered as due to photoionization and the
estimation of the same effect for HeII (Jakobsen et al. 1994; Tytler
et al. 1995; Davidsen et al. 1996).

We will calculate the He ionization fractions for absorbers of given
density due to UV fluxes corresponding to the different alternatives to
inspect their agreement with the observations of LLS. Additionally, from the expressions
of GP effect for HI and HeII we will obtain a bound for the ionization
fraction of the latter in IGM to compare its consistence with that of
Lyman-$\alpha$ clouds (LC) for the various possibilities of flux.

In Section 2 we will describe the possible sources of UV radiation whose
difference is roughly that QSOs ionize HI, HeI and HeII, stars just HI and
HeI and decaying neutrinos only HI. The contribution of stars will be fixed
by recent determinations of the proximity effect
(Giallongo et al. 1996) and that of DDM requiring that it is capable to ionize
alone the HI and NI of the Milky Way (Sciama 1990b) or that the decay photons
have not enough energy to ionize NI (Sciama 1995).

In Section 3 the numerical calculation will be presented for absorbers of different
density with the alternative UV fluxes to determine the ionization of He and other
properties and see their compatibility with
 observations in LLS.

Section 4 will be devoted to compare GP effect for HI and HeII with the
different models estimating the fraction of the latter in homogeneous IGM to
evaluate through approximate formulae if it is consistent with the
numerically calculated properties of LC.

The conclusions will be given in Section 5 indicating the possible relation
of the UV ionizing flux with different cosmological models.

\vspace{0.3cm}
\noindent { \bf 2. Sources of UV background }
\vspace{0.2cm}

Recent determinations of bounds  for GP effect for
HI and estimations for HeII give, under the assumption of
photoionization to explain the ratio of the optical depths

\begin{equation}
\tau ^{GP}_{HeII} / \tau^{GP}_{HI} = 0.45\; J_{HI} / J_{HeII} \,\, ,
\end{equation}
a ratio of the effective UV fluxes at the corresponding ionization frequencies of at
least

\begin{equation}
S_{L} = J_{HI} / J_{HeII} = 100
\end{equation}
 for $z = 3.3$. Even if
the GP effect is masked by line blanketing, the above ratio is
maintained rather high with the estimation $S_L \geq 40$ (Madau \& Meiksin
1994) or $S_{L} \geq 65 $ (Sethi 1995). On the other hand, from metallicity
abundance it is found that $S_L \sim 70$ at $z \sim 3.2$ (Songaila et al.
1995) and even $S_L > 100$ for $z = 3.5 - 3.8$ (Savaglio et al. 1996).

This ratio $S_{L}$ seems to exceed that due to QSOs which was
evaluated as $\sim 30$
(Madau \& Meiksin 1994; Haardt \& Madau 1996) in agreement with the
declining quasar population for $ z > 3$ (Pei 1995). It appears therefore
necessary to fill the difference with another UV source.

If one adds stars of primeval galaxies
(Miralda-Escud\'e \& Ostriker 1990) they may ionize HI and HeI with a
spectrum similar to that of QSOs but which afterwards drops
abruptly so that HeII is ionized only by the latter source.
 If one fits the resulting flux of QSOs in
the range between HI and HeI  as $J \sim
\nu^{-\alpha}$ with $\alpha \leq 1$, it follows that for the total flux

\begin{equation}
J_{HI} / J_{HeI} \leq 2 \, \, .
\end{equation}

Another possibility is the addition to QSOs of DDM thought (Sciama 1993)
as neutrinos of mass around 30 eV which could correspond to close the
universe with HDM. This model is alternative to the first since HDM
delays galaxy formation so that stars would be excluded. According to the
details of the decaying neutrinos the relation of Eq.~3 may be altered.

Our analysis will be based on considering three models for UV fluxes:
that due to QSOs alone and the alternatives of adding either stars or
decaying neutrinos. We will consider them at $z = 2$ and $4$ because
the former redshift corresponds to the observations of LLS, and the latter to
the range of GP estimations for HI and HeII (see Fig. 1).

\begin{figure}
\caption[1]{\label{f1} The UV ionizing source as function of the frequency
(in Rydberg) for two different redshifts.
The continuum line is for QSO sources only, the dashed line is
the same with the addition of stars, the dotted line is with the addition of
decaying dark matter (DDM) photons. The upper two plots represent the
UV flux in case of strong contribution by the QSOs (sQSO), and in case of
the QSO+DDM model, of strong contribution by the DDM (sQSO+sDDM).
The lower two panels are
for weak QSO contribution (wQSO) and, in the case of the DDM model, for
strong QSO contribution, but weak DDM contribution (sQSO+wDDM).}
\epsfxsize=13cm
\epsfysize=13cm
\centerline{\epsffile{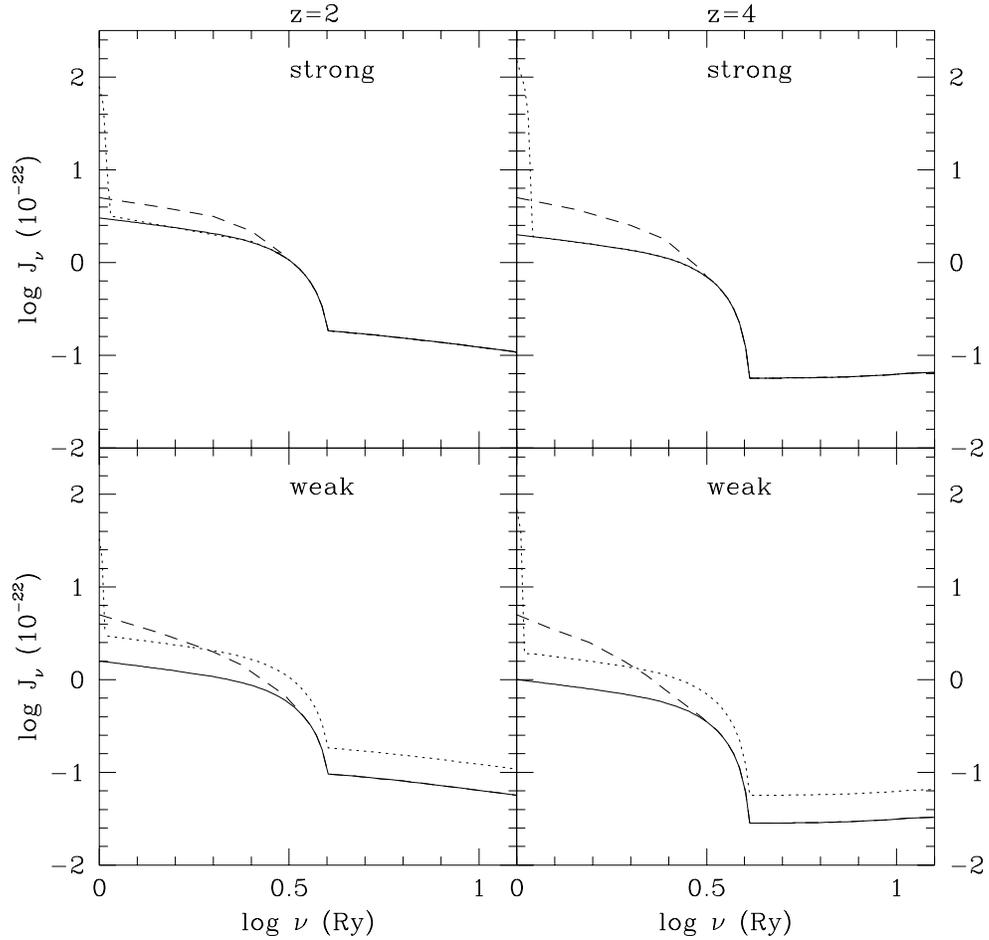}}
\end{figure}

 Using for the effective UV flux the
 normalization\\ $J = J_{-22} \times 10^{-22}$ erg s$^{-1}$
cm$^{-2}$ ~Hz$^{-1}$ sr$^{-1}$,
the evaluation of the QSO contribution has increased doubling the
earlier values (Madau 1992) to reach at $1 Ry$ a maximum of $J_{-22} \sim
3$ for $z \geq 2$ when dust obscuration is taken into account (Haardt \&
Madau 1995) and diminishing for larger redshifts so that
 we estimate $J_{-22} \sim 2$ at $z \sim 4$.
According to the discussed values of $S_L$ due to QSOs, and considering that
this ratio increases with $z$, one evaluates as maximum contribution at
 $4 Ry \,\,\,\, J_{-22}
\sim 0.05$ at $z = 4$ and slightly above $0.1$ at $z = 2$, which will
coincide with the total flux at this frequency in all the alternative
models. We will denote this scheme as the strong QSO model (sQSO).

Stars give a slightly softer spectrum in the range HI -- HeI and we normalize
the total flux in the QSO+star model at $1 Ry$ as $J_{-22} \sim 5$ to agree
with recent determinations from proximity effect  $J_{-22} = 5 \pm 1$ at
 $1 Ry$ which show no evolution in the range $2 < z < 4$ (Giallongo et al.
 1996). In this way it will be noted
that the ratio of Eq.~2 is satisfied for $z \sim 4$ and that of
Eq.~3 reasonably fulfilled in the range $2 < z < 4$ . To explore the
situation $S_L = 200$ at $z \sim 4$ we will also consider the weak QSO
contribution (wQSO) corresponding to take $50\%$ of the above quoted values,
 always keeping the normalization $J_{-22} = 5$ at $1 Ry$ for the total
 QSO+star flux.

It must be remarked that the fluxes denoted as $J_{HI}$, $J_{HeI}$ and
$J_{HeII}$ are the average over frequency weighted with the corresponding
ionization cross section. In the QSO
and QSO+star cases they differ only slightly from the flux $J_{\nu}$ at
$\nu = 1 Ry$ etc, because of the fast decrease of the cross section. E.g.

\begin{equation}
J_{HI} = \frac {\int_{1 Ry}^{\infty} J_\nu \, (\sigma_{HI}/\nu) \, d\nu}
	   {\int_{1Ry}^{\infty} (\sigma_{HI}/\nu) \, d\nu} \,\,\, .
\end{equation}

On the other hand for the QSO+DDM model the only difference from QSOs alone
is a large peak around $1 Ry$ so that this maximum of $J_\nu$  will be
much greater than the averaged $J_{HI}$. From the requirement
(Sciama 1990b) that decaying neutrinos are able to ionize the H of
the Milky Way their lifetime must be $\tau_\nu \sim 2 \times 10^{23}$ s.
Correspondingly, the intergalactic flux expressed in number of photons per
${\rm cm}^2$ and s at $z \sim 0$ would be

\begin{equation}
F = \frac{n_\nu}{\tau_\nu} \frac{c}{H_0} \frac{\epsilon}{13.6}
\end{equation}
with photon energy $E_\gamma = (13.6+ \epsilon)$ eV, and where $n_\nu$ is
density of neutrinos and $H_0$ present Hubble constant. Assuming (Sciama 1993)
that decay photons are able to ionize NI, the neutrino mass must be
 $m_\nu \sim 29$ eV  and consequently
$F \sim 6 \times 10^5$ which corresponds to $J_{-22} \sim 24$.
Since according to this model the flux due to neutrinos increases with
redshift as $(1+z)^{3/2}$, we would have $J_{-22} \sim 120$ at $z=2$ .

The ionization cross-section has the frequency behaviour $\sigma \sim
\nu^{-3}$ so that for a peak of width $\epsilon$ additional to the spectrum
due to QSOs one has, according to Eq.~4,

\begin{equation}
J_{HI} = J^{DDM} \frac{3~\epsilon}{13.6} + J_{1 Ry}^{QSO} \frac{3}{3+\alpha}
			 \,\,\, .
\end{equation}
For the above values, using strong QSO contribution, one would obtain \\
$J_{HI,-22} \sim 27$ at $z = 2$ which is extremely large. Considering a reduction
of the flux due to absorption by clouds, one may estimate at
$1 Ry$ $J_{-22} \sim 80$ for $z = 2$, case which we will denote
as strong DDM model (sDDM). As a consequence, the
averaged flux for QSO+DDM will be $J_{HI,-22} \sim 18$ and we may
establish a typical ratio (Sciama 1994) at $z = 2$ for this standard
QSO+DDM model

\begin{equation}
J_{HI} / J_{HeI} \sim 8 \,\, .
\end{equation}

For $z = 4$ the same strong DDM contribution will be $J_{-22} \sim
 150$ which gives an averaged flux $J_{HI,-22} \sim
 30$ for sQSO+DDM and therefore a ratio $S_L$ much larger than that of Eq.~2.
It is seen that the flux due to
neutrinos is higher than the large initial observations of proximity effect at $z \sim
2.5$ (Bajtlik et al. 1988) and even beyond the upper quoted values (Bechtold
1994). Other determinations give smaller fluxes which either would be consistent
with QSOs alone (Williger et al. 1994; Lu et al. 1996) or correspond to
 the value we have taken to
normalize the QSO+star model (Giallongo et al. 1996). Therefore for the
described QSO+DDM case a non-standard interpretation of the proximity effect
 must be invoked (Sciama 1991).

But recently a modified decaying neutrino has been proposed (Sciama 1995) due
to an estimation of the metagalactic flux $F$ at $z \sim 0$ four times smaller
than that given by $m_\nu \sim 29$ eV. Therefore, keeping the value of the
lifetime, from Eq.~5 $\epsilon$ must not exceed $0.2$  and the decay
photons cannot ionize NI. The maximum fluxes due to this weak DDM (wDDM)
would be $J_{-22} \sim 30$ at $z = 2$ and $J_{-22} \sim 67$ at $z = 4$ which
would be roughly compatible with the highest estimations from proximity effect.
Due to the fact that the peak is now smaller and narrower, from Eq.~6 the averaged flux
 for the modified DDM+sQSO model turns out to be $J_{HI,-22} \sim 4.4$ at
$z = 2$ and $J_{HI,-22} \sim 5$ at $z = 4$ . Therefore the bound (3) is
almost respected and that of Eq.~2 exactly satisfied. We note that we have
performed a more accurate relation between neutrino mass and flux of decay
photons than that done in Miralda-Escud\'e \& Ostriker 1992.

 Since the standard QSO+sDDM model satisfies the Eq.~7 whereas the QSO+star
alternative as well as QSOs alone and the modified QSO+wDDM model correspond
 to a ratio (3), one may test their consistence using them in the ratio of
 neutral
densities which follows from photoionization equilibrium for HI and HeI
(Miralda-Escud\'e \& Ostriker 1992; Sciama 1994)

 \begin{equation}
\frac{n_{HeI}}{n_{HI}} = 0.044 \frac{J_{HI}}{J_{HeI}} \chi_{HeII} \,\,\, .
\end{equation}
 $\chi_{HeII}$ is the fraction of single-ionized He
and the numerical coefficient is independent of temperature and density of
the system. Moreover, information from GP effect which involves HI
and HeII does not support QSOs as the only UV source and may distinguish between
 QSO+star model and the modified QSO+wDDM possibility.

\vspace{.3cm}
\noindent 3. {\bf  Lyman-limit systems and Lyman-$\alpha$ clouds}
\vspace{.2cm}

For LLS a ratio $n_{HeI} / n_{HI} \sim 1/30$ has been observed (Reimers \&
Vogel 1993) for $z\simeq 2$. If one takes the standard QSO+sDDM alternative
 of Eq.~7, Eq.~5 would require that for $z\simeq 2 \,\,\,\,\,
\chi_{HeII} \simeq 0.1$ the rest being almost all HeIII. On the contrary
if one chooses either the fluxes of QSOs alone or those of QSO+star or modified
QSO+wDDM models, it follows from Eq.~3 that $\chi_{HeII} \geq 0.5$.

To obtain the fractions of ionized
He, we have performed a numerical calculation for absorbers using the standard
photoionization code CLOUDY (Ferland 1991).

As input for the LLS we have taken
 the HI column density \\ $N_{HI} \sim 10^{17}$ cm$^{-2}$, the metallicity as
$1/100$ of the solar value and the H density $n_H \sim 10^{-2}$ cm$^{-3}$.
This last figure corresponds to the lower bound (Steidel 1990; Lanzetta 1991)
 consistent with sizes smaller than $15$ kpc.

 For the LC we have fixed $N_{HI} \sim 10^{14}$ ${\rm cm}^{-2}$, $n_H \sim 10^{-4}$
${\rm cm}^{-3}$  and metallicity 1/1000 of the solar value,
which correspond to their commonly accepted sizes (Meiksin \& Madau 1993). Lower limits to the transverse size of the LC observed in
the gravitational lensed spectra of HE1104-1805 (Smette et al. 1995) is of
the order of $50~h^{-1}$ kpc at $2 \sigma$ level for spheroidal clouds at
$z \sim 2$, where  $h = H_0 / (100$ km s$^{-1}$ Mpc$^{-1})$ . Similar results have been found (Dinshaw et al. 1995) from the
cross correlation  of absorption lines in the close quasar pair 1343+2640A and
B deriving sizes larger than $40~h^{-1}$ kpc at $z= 1.8$.

We have assumed that the UV background radiation
is the only ionizing source and we used for the flux
the different shapes and intensities
according to Fig.~1. Results are summarized in Tables 1 and 2.

\begin{table}
\caption[t1]{Fraction of HeII and HeIII for different UV background models
(as shown in Fig. 1) in the LLS.}
\begin{center}
\begin{tabular}
{l|ccc|ccc}
\hline&&&&&&\\[-5pt]
\multicolumn{1}{c|}{Model}
& \multicolumn{3}{c|}{$z=2$}
& \multicolumn{3}{c}{$z=4$} \\
&$J(1Ry)_{-22}$ & $\chi_{HeII}$ & $\chi_{HeIII}$ &$J(1Ry)_{-22}$ & $\chi_{HeII}$ & $\chi_{HeIII}$ \\
\hline&&&&&&\\[-5pt]
sQSO      & 3   & 0.60 & 0.40 & 2   & 0.78  &  0.22 \\
wQSO      & 1.6 & 0.70 & 0.29 & 1   & 0.86  &  0.14 \\
sQSO+star & 5   & 0.67 & 0.33 & 5   & 0.84  &  0.15 \\
wQSO+star & 5   & 0.79 & 0.20 & 5   & 0.91  &  0.09 \\
sQSO+sDDM & 80  & 0.79 & 0.21 & 150 & 0.94  &  0.06 \\
sQSO+wDDM & 33  & 0.67 & 0.33 & 69  & 0.87  &  0.12 \\
[4pt]\hline\end{tabular}\end{center}

\caption[t2]{Fraction of HeII and HeIII for different UV background models
(as shown in Fig. 1) in the LC.}
\begin{center}
\begin{tabular}
{l|ccc|ccc}
\hline&&&&&&\\[-5pt]
\multicolumn{1}{c|}{Model}
& \multicolumn{3}{c|}{$z=2$}
& \multicolumn{3}{c}{$z=4$} \\
&$J(1Ry)_{-22}$ & $\chi_{HeII}$ & $\chi_{HeIII}$ & $J(1Ry)_{-22}$ & $\chi_{HeII}$ & $\chi_{HeIII}$ \\
\hline&&&&&&\\[-5pt]
sQSO      & 3   & 0.004  & 0.996 & 2   & 0.013  & 0.987 \\
wQSO      & 1.6 & 0.009  & 0.991 & 1   & 0.029  & 0.971 \\
sQSO+star & 5   & 0.004  & 0.996 & 5   & 0.013  & 0.987 \\
wQSO+star & 5   & 0.008  & 0.992 & 5   & 0.026  & 0.974 \\
sQSO+sDDM & 80  & 0.004  & 0.996 & 150 & 0.015  & 0.985 \\
sQSO+wDDM & 33  & 0.004  & 0.996 & 69  & 0.014  & 0.986 \\
[4pt]\hline\end{tabular}\end{center}
\end{table}

Considering LLS, it turns out that at $z = 2 \,\,\,\, \chi_{HeII}
> 0.5$ for the three models. In particular for the modified decaying neutrino in addition to the sQSO contribution,
 the values of $\chi_{HeII}$ turn out to be similar to those of the
sQSO+star model. Therefore $\chi_{HeII} \sim 0.1$ is excluded with all the
 models for the chosen value of the density $n_H$.

One must remark that the observation of $n_{HeI}/n_{HI}$ in LLS was accompanied
by that of ionized states of C, N and O (Reimers \& Vogel 1993) which would be
consistent with a large He ionization, i.e. $\chi_{HeII} \sim 0.1$. This led to
 the explanation of the whole set of observations with the standard decaying
neutrino (Sciama 1994) using the large ratio (7) of $J_{HI}/J_{HeI}$. But to
obtain $\chi_{HeII} \sim 0.1$ it is necessary to decrease the density $n_H$ in
such a way that the sizes of LLS would be larger than $100$ kpc which is an
unattractive possibility. This has suggested that the heavy elements are
collisionally ionized in hot regions (Giroux et al. 1994) different from those
where H and He are photoionized. This is the scenario adopted for our analysis.

Therefore we may state that QSO+star model, QSOs alone and also the modified
sQSO+wDDM alternative are consistent with the observation of $n_{HeI}/n_{HI}$
at $z = 2$. From the tables we may see that the CLOUDY calculation predicts
a smaller He ionization at $z = 4$ as it is generally expected.

For further considerations it is convenient to use approximate
expressions assuming photoionization equilibrium for HI and HeII in a
thin absorber of size $R$ expressed in kpc and column density given in
cm$^{-2}$, which allow to write (Miralda-Escud\'e \& Ostriker 1992)

\begin{equation}
\chi_{HeII} = \left(\frac{N_{HI} \,\, J_{HI,-22}}{R} \right)^{1/2}
		\frac{2.7 \times 10^{-10}}{J_{HeII,-22}} \,\,\, ,
\end{equation}
where the numerical figure includes the square root of the recombination
coefficient whose temperature dependence is (Spitzer 1978)  $\alpha (T) \sim
T^{-3/4}$ and, defining \\ $T = T_4~10^4~K$, $T_4 \sim 2$ has been taken.
 It is possible to check the densities for which
the approximate expressions are valid so that also IGM can be included
 in the comparison.

For a thicker absorber like LLS the left-hand side of Eq.~9 must be
replaced in principle by $\chi_{HeII} / (1-\chi_{HeII})$ since the
right-hand side gives $n_{HeII} / n_{HeIII}$. But in this way the obtained
$\chi_{HeII}$ will be a lower bound because self-shielding for He
should be already included for LLS, diminishing the value of $J_{HeII}$. Another
source of approximation of Eq.~9 is the fact that instead of the
detailed flux spectrum, the averaged values are included what is particularly
relevant in $J_{HI}$ for the QSO+DDM case, as indicated in the previous
Section.

Comparing with our more accurate results of Table 1 for LLS, it is easily
seen that the approximate formula gives typically a value of $\chi_{HeII}$
lower in $0.1$ for QSOs and a slightly larger difference for QSO+star and
QSO+DDM models. In fact for $z = 2$ CLOUDY gives e.g. for sQSO, sQSO+star,
sQSO+sDDM and sQSO+wDDM models (Table 1)
sizes $R \simeq 1.4,~2.4,~5.2$ and $2.2$ kpc respectively, with a temperature
$T_4 \sim 2$. The difference
 can be understood if in the approximate formula $J_{HeII}$ is
reduced around $40\%$ in the QSO case because of self-shielding, around
$50\%$ in the QSO+star and QSO+wDDM models and even more than $60\%$ for
 QSO+sDDM due to the
increasing size of the absorber which results from CLOUDY.

For thin absorbers like LC, one expects that both CLOUDY
and the approximate Eq.~9 give similar results because
 the absorber becomes thick to $J_{HeII}$ only at
$N_{HI} \sim 10^{15}$ cm$^{-2}$ (Miralda-Escud\'e \& Ostriker
1990). In fact CLOUDY gives for LC a
ionization larger than for LLS with a typical $\chi_{HeII} \geq 0.01$ at $z = 4$
(Table 2).
One may check that Eq.~9 gives the same values of $\chi_{HeII}$
once the numerical coefficient is changed in correspondence with the
different $T$. For this comparison we must quote that CLOUDY
for $z = 4$ gives for sQSO, sQSO+star, sQSO+sDDM and sQSO+wDDM models $R \simeq 17,~
37,~160$ and $47$ kpc respectively with $T_4 \sim 4.5$.

CLOUDY has produced rather large temperatures for LC which might result
from thermal evolution of collapsed systems (Miralda-Escud\'e \& Rees 1994)
and that correspond to Doppler
broadening in the range 25 \kms~$\lsim b \lsim 32$ \kms, which is marginally
consistent with the measured values obtained from high resolution QSO spectra.
 This would require gravitational confinement of LC in a colder IGM.

Considering only the photoionization equilibrium for HeII, it is also
possible to check the consistency of predictions for different thin
absorbers through

\begin{equation}
\frac{\chi_{HeII}}{1 - \chi_{HeII}} = \frac{\alpha(T) \, n_{e}}{J_{HeII} \,
\sigma_{HeII}} \,\,\, .
\end{equation}
Eq.~10 is one of the ingredients which lead to Eq.~9 and involves only
the density and temperature of the medium so that it is valid
also for homogeneous IGM.

Since for LLS and LC CLOUDY gave the results of Tables
1 and 2, it is clear that the discrepancies in the
use of Eq.~10 to compare these absorbers must come from self-shielding
effects in LLS. Thinking in principle that UV flux and ionization cross
section are common properties, the ratio of Eq.~10 for LLS and LC
should be

\begin{equation}
{[\chi_{HeII} / \chi_{HeIII}]}_{LLS} \,\, {[\chi_{HeIII} /
\chi_{HeII}]}_{LC} = (n_{H_{LLS}} / n_{H_{LC}}) \,\, (T_{LC} /
T_{LLS})^{3/4} \,\,\, .  \end{equation}

If we take the values of Tables 1 and 2 for QSO, QSO+star and QSO+DDM models at $z
= 2, \, 4$ being typically $T_4 \sim 2$ for LLS and $5$ for LC,
the left-hand side of Eq.~11 is larger than the right-hand side
in an amount which is explained by the same reduction of $J_{HeII}$ for
LLS that we discussed in connection with the discrepancies of Eq.~9.

Therefore we conclude that the approximate expressions based on
photoionization equilibrium work well for thin absorbers so that we may use
them to compare LC with the less dense homogeneous IGM.

\vspace{.3cm}
\noindent 4. {\bf Homogeneous Intergalactic Medium}
\vspace{0.2 cm}

We anticipate that the results of this Section cannot be too precise due
to the large uncertainties in the properties of homogeneous IGM.

As a first step we
consider GP effect for HI which, if the ionization is due entirely to UV
radiation, will correspond to the optical depth

\begin{equation}
\tau^{GP}_{HI} = 23  \left(\frac{\Omega_{IGM} h^2} {0.015} \right)^2
		      \left( \frac{0.5}{h} \right)
		     \left( \frac{1+z}{5} \right)^6
		     \frac{H_0}{H(z)}
		     \frac{T_4^{-0.7}}{J_{HI,-22}}
\end{equation}
where $\Omega_{IGM}$ is the fraction of critical density in the homogeneous
IGM.

>From the bound (Giallongo et al. 1994) at $z = 4.3 \,\,\,\,\, \tau_{HI}^{GP} \leq
0.02$ we will obtain the bounds on density of IGM for different UV models
which then, through GP for HeII, will allow to estimate $\chi_{HeII}$ .

We start with the sQSO+star model with $S_L = 100$ at $z = 4$, taking $T_{4}
\sim 2$ which corresponds to inhomogeneous photoionization
(Miralda-Escud\'e \& Rees 1994). We adopt the value of the Hubble
constant $h = 0.7$ which seems to emerge recently (Freedman et al. 1994;
Riess et al. 1995; Tanvir et al. 1995), though also smaller values have been
obtained (Tammann et al. 1996; Branch et al. 1996), that suggests the presence of the
cosmological constant (Krauss \& Turner 1995). Since for \\ $\Omega_{\Lambda}
\sim 0.6$ plus CDM, model denoted as $\Lambda$CDM, the ratio $\,\, H(4)/H_{0} \simeq 7.1$ is obtained (Reisenegger \&
 Miralda-Escud\'e 1995), the estimation  $\Omega_{IGM} \leq \frac{1}{5}
\Omega_B$ follows from the bound of Eq.~12 where we have taken the average
of the quoted values of the baryonic density $\Omega_B = 0.015 / h^2$. It is
interesting that this agrees with simulations for $\Lambda$CDM giving $80\%$
of baryons in collapsed form (Miralda-Escud\'e et al. 1995).

We now pass to GP for HeII whose optical depth is

\begin{equation}
\tau^{GP}_{HeII} = 2.5  \times 10^5
\left( \frac{\Omega_{IGM} h^2}{0.015} \right)
\left (\frac{0.5}{h} \right)
\left (\frac{1+z}{4}\right)^3
\frac{H_0}{H(z)}
y_{HeII}
\end{equation}
where $y_{HeII} = n_{HeII} / n_H$. The observations (Jakobsen et al. 1994;
 Tytler et al. 1995; Davidsen et al. 1996) and their interpretation
 (Shapiro 1995) allow to estimate \\ $\tau^{GP}_{HII}\sim 1$ at $z = 3.3 $ \,
.

With the above parameters and saturating the bound for $\Omega_{IGM}$ the
QSO+star model gives from Eq.~13 $y_{HeII} \simeq 1.7 \times
10^{-4}$ and, using the cosmological ratio between H and He, $\chi_{HeII}
 \simeq 2 \times 10^{-3}$. A similar prediction results from the
addition of modified decaying neutrinos to QSOs. For QSOs alone the flux $J_{HI,-22} \sim 2$ at
 $z \sim 4$ would produce instead a larger fraction $\chi_{HeII} \simeq 3 \times 10^{-3}$.

For the standard QSO+sDDM alternative if one takes the averaged flux \\ $J_{HI,-22} \sim
30$ at $z = 4$, $T_{4} \sim 1$ due to the lower temperature of homogeneous
photoionization and the rest of parameters as required by the decaying neutrino
model (Sciama 1990a) $ \Omega = 1, \,\, H(z)/H_0 = (1 +
z)^{3/2}$ ,  $h = 0.56$ (which would coincide with the measurement of
Tammann et al. 1996 and Branch et al. 1996), from Eq.~12 it turns out
 $\Omega_{IGM} \leq \frac{2}{5} \Omega_B$. The saturation of this
bound for $\Omega_{IGM}$ would be in agreement with the delayed formation of
structures predicted by HDM which requires $\Omega_{IGM} \geq \frac{1}{3}
\Omega_B$  (Williger et al. 1994). In  this way one would obtain from Eq.~13
 $y_{HeII} \simeq 0.9 \times 10^{-4}$
 and  $\chi_{HeII} \simeq 1.1 \times 10^{-3}$.

We now compare these results for IGM with those for LC of Table 2 using
Eq.~10 from which

\begin{equation}
\frac {\chi_{HeII_{LC}}}{\chi_{HeII_{IGM}}} = \frac
{n_{H_{LC}}}{n_{H_{IGM}}} \,\, \left(\frac {T_{IGM}}{T_{LC}}\right)^{3/4}
\,\,\, .  \end{equation}

Taking as average the properties of IGM at $z \sim 4$ it seems clear
that the spectrum of QSOs alone is excluded because, since  $n_{H_{IGM}}
\sim 10^{-5} \, \Omega_{IGM}/\Omega_{B} \, {\rm cm}^{-3}$ (Madau \& Meiksin
1994), inserting the rest of parameters in Eq.~14 the left-hand side
is one order of magnitude smaller than the right-hand side.  This conclusion
is consistent with the fact that the UV spectrum must be softer than that of
QSOs to agree with the estimations of GP as seen in Eq.~1.

With the QSO+star model the left-hand side of Eq.~14 increases
because $\chi_{HeII_{IGM}}$ is smaller, and the right-hand side decreases
because $n_{H_{IGM}}$ is larger so that the disagreement diminishes.
A similar situation applies to the addition of modified decaying neutrinos
to QSOs. One must note that this conclusion is independent of the chosen value of $h \,
H(z) / H_0$ since both $\chi_{HeII}$ and $n_H$ for IGM turn out to depend
on the square root of it.

For the sQSO+sDDM alternative the comparison of both sides of Eq.~14 is
reverted because, to the
further modifications of $\chi_{HeII}$ and $n_H$ for IGM, one must add its
predicted smaller temperature so that the left-hand side becomes almost twice
larger than the right-hand side. This indicates that $S_L$ may be too large in
this case.

It is interesting to note that for the QSO+star model the increase of $S_L$
should improve the agreement of Eq.~14. In fact with our reduction of $50\%$
of the flux due to QSOs which gives $S_L = 200$ at $z = 4, \,\,\,\, \chi_{HeII}$ for IGM increases
in $100\%$ because one would expect $\tau_{HeII}^{GP} \sim 2$. But looking at
Eq.~9 $\chi_{HeII}$ for LC increases more because, apart from the influence
of $J_{HeII}$, a smaller global flux decreases $R$ and $T$ giving a further
enhancing factor. However the improvement is not fast for $S_L = 200$
as seen in Table 2.

We must note that the same procedure of using a weak QSO contribution cannot
be applied to the modified decaying neutrino because, in doing so, one would
obtain a ratio $J_{HI}/J_{HeI}$ at $z = 2$ which would clearly exceed the bound
of Eq.~3. Therefore the only way to improve the comparison of the QSO+wDDM
model with homogeneous IGM would be to accept a decline of QSO contribution
for $z > 3$ more marked that the generally estimated $exp[-0.69 (z-3)]$ assumed
here.

Apart from the modification of decaying neutrinos obtained diminishing slightly
 its mass, one could keep this value to allow NI ionization and increase its
 lifetime.
Relaxing the condition that neutrinos ionize alone the HI in the
Milky Way (Dodelson \& Jubas 1994) one might assume the  $\nu$
to be more stable in one order of magnitude $i.e.$   $\tau_{\nu} \sim
10^{24}$ s.

Several models for massive decaying neutrinos give the dependence of
 lifetime on mass and magnetic moment

\begin{equation}
\tau_{\nu} \simeq \left(\frac {29 \, eV}{m_{\nu}}\right)^{3} \,\,
\left(\frac {10^{-14} \, \mu_{B}}{\mu}\right)^{2} \,\,\, 0.8 \times 10^{23}
\, s \end{equation}
and in particular the minimal supersymmetric standard model with couplings
which violate the $R$ parity allows easily (Roulet \& Tommasini 1991)
$m_{\nu} \sim 29$ eV and \\ $\mu \sim 10^{-14} \, \mu_{B}$ . To increase the
lifetime to $\tau_{\nu} \sim 10^{24}$ s one must decrease slightly these
coupling constants and, to keep the value of $m_{\nu}$ , correspondingly
increase the scale of supersymmetric partners above $100$ GeV. This is
perfectly possible in the range of the theoretically admissible parameters.

The averaged DDM flux would be similar to that of the wDDM previously
described so that again, added to that of sQSO model, the properties of LLS
would be reproduced but the consistence between LC and IGM would be difficult
because of not high enough $S_L$.

It seems that the most convenient shape of UV spectrum for a general agreement
is one where the flux decreases gently between HI and HeI ionization
frequencies and then drops abruptly towards the HeII edge. A modification of
DDM that might give this result is to further increase $\tau_{\nu}$ above
$10^{24}$ s and simultaneously double the neutrino mass to allow the
 ionization of both HI and HeI, which would be admissible if $h>0.74$ (Bradford
\& Hogan 1996), but since this implies a major change of the model we will not
analyze in detail its consequences here.

Obviously all our discussion related to homogeneous IGM is based on assuming
photoionization as seen in Eq.~12. Whereas for densities $n \sim
10^{-2} \, - \, 10^{-4}$ cm$^{-3}$ collisional ionization is not
important (Haardt \& Madau 1995), it may be relevant in low density IGM if the
temperature is higher than what assumed here. In this case the conclusions
regarding UV sources would not apply to IGM.

\vspace{.3cm}
\noindent 5. {\bf Conclusions}
\vspace{.2cm}

Using a photoionization code for absorbers we have seen that the
QSO+star model for UV sources is able to reproduce the bulk of properties of
homogeneous IGM, LC and LLS, requiring that $S_L > 100$ at $z \sim 4$ in
agreement with what emerges from recently observed metallicity abundance.
 For the comparison
with IGM it would be crucial to determine its temperature with more
precision. These same properties are not easily reconciled assuming the
addition to QSOs of neutrinos of mass $\sim 29$ eV which would complete
the critical Universe mass with only HDM. In fact if one takes the lifetime
$\tau_{\nu} \simeq 2 \times 10^{23}$ s, there is a rough agreement between LC
and IGM but properties of LLS are not reproduced. On the other hand if one
reduces  the neutrino mass to $27.6$ eV or increases its lifetime in one
order of magnitude, the LLS difficulty is solved but the matching of LC and
IGM properties becomes questionable unless there is a very fast decline of QSOs
for $z > 3$.

It is interesting that the original model of decaying neutrinos
gives, through GP bounds, the large density of non-collapsed baryonic
matter predicted by HDM. In the same way the  QSO+star model is
consistent with the scenario of CDM  with cosmological constant
$\Omega_{\Lambda} = 0.6$ convenient to explain the X-ray emitting
intracluster gas, as well as old galaxies at large redshift (Krauss 1996),
 but in difficulty with recent measurement of the
deceleration parameter (Dodelson et al. 1996). Therefore it seems that when
GP effect and properties of absorbers will be better determined, the
observed properties of the reionization of Universe will give definite
hints, regarding the models which originated the structures, at a redshift
intermediate between the one of recombination age and that of formation of bulk of
galaxies.

\vspace {.5 cm}
We are deeply indebted to Prof. D.W. Sciama for comments related to modified
decaying neutrinos.
S.S. thanks heartily the hospitality at the Osservatorio Astronomico di Roma
where part of this research was performed. L.M. acknowledges
partial financial support from CONICET of Argentina through grant PID
339650092.

\vspace {.8 cm}
\noindent {\bf References}

\vspace{.2cm}
\noindent Bajtlik S., Duncan R.C. \& Ostriker J.P., 1988, {\it ApJ} {\bf 327}, 570.
\\Bechtold J., 1994, {\it ApJS} {\bf 91}, 1.\\
Bradford E. \& Hogan C., 1996, ASTRO-PH 960475, {\it ApJ} (in press).\\
Branch D., Fischer A., Baron E. \& Nugent P., 1996, ASTRO-PH 9604006.\\
Davidsen A.F., Kriss G.A. \& Zheng W., 1996, {\it Nature} {\bf 380}, 47.\\
Dinshaw N., Impey C.D., Foltz C.B., Weymann R.J. \& Morris S.L., 1995, {\it
 Nature}\\ \indent {\bf 373}, 223.\\
 Dodelson S. \& Jubas J.M., 1994, {\it MNRAS} {\bf 266}, 886.\\
Dodelson S., Gates E.I. \& Turner M.S., 1996, ASTRO-PH 9603081 (submitted to
\\ \indent {\it Science}).\\
Ferland G.J., 1991, {\it OSU Astronomy Dept. Internal Rept.} 91-01.\\
Freedman W.L., Madore B.F., Mould J.R., Hill R.,
Ferrarese L., Kennicutt R.C.,\\ \indent Saha A., Stetson P.B., Graham J.A.,
 Ford H., Hoessel J.G., Huchra J.,\\ \indent Hughes S.M.  \&
 Illingworth G.D., 1994, {\it Nature} {\bf 371}, 757.\\
 Giallongo E., D'Odorico S., Fontana A.,
McMahon R., Savaglio S., Cristiani S.,\\ \indent Molaro P. \& Trevese D.,
1994, {\it ApJ} {\bf 425}, L1.\\
Giallongo E., Cristiani S., D'Odorico S., Fontana A. \& Savaglio S., 1996,\\
\indent {\it ApJ} (in press).\\
Giroux M.L., Sutherland R.S. \& Shull J.M., 1996, {\it ApJ} {\bf 435}, L97.\\
 Gunn J.E. \& Peterson B.A., 1965, {\it ApJ} {\bf
142}, 1633.\\
 Haardt F. \& Madau P., 1996, {\it ApJ} {\bf 461}, 20.\\
 Jakobsen P., Boksenberg A., Deharveng J.M., Greenfield P., Jedrzejewski R.  \&\\ \indent
Paresce F., 1994, {\it Nature} {\bf 370}, 35.\\
Krauss L.M., 1996, ASTRO-PH 9607103.\\
 Krauss L.M. \& Turner M.S.,
1995, {\it J. Gen. Rel. Grav.} {\bf 27}, 1137.\\
Lanzetta K.M., 1991, {\it ApJ} {\bf 375}, 1.\\
Lu L., Sargent W.L.W., Womble D.S. \& Takada-Hidai M., 1996, ASTRO-PH 9606033,\\
\indent {\it ApJ} (in press).\\
 Madau P., 1992, {\it ApJ} {\bf 389},
L1.\\ Madau P. \& Meiksin A., 1994, {\it ApJ} {\bf 433}, L53.\\
Meiksin A. \& Madau P., 1993, {\it ApJ} {\bf 412}, 34.
\\ Miralda-Escud\'e J. \&
Ostriker J.P., 1990, {\it ApJ} {\bf 350}, 11.
\\ Miralda-Escud\'e J. \& Ostriker J.P., 1992, {\it ApJ} {\bf 392}, 15.\\
 Miralda-Escud\'e J. \& Rees M.J., 1994, {\it MNRAS} {\bf 266}, 343.\\
 Miralda-Escud\'e J., Cen R.,
Ostriker J.P. \& Rauch M., 1995, ASTRO-PH 9511013.\\
Pei Y.C., 1995, {\it ApJ} {\bf 438}, 623.\\
Reimers D. \& Vogel S.,
1993, A\&A {\bf 276}, L13.\\ Reisenegger A. \& Miralda-Escud\'e J., 1995,
{\it ApJ} {\bf 449}, 476.\\ Riess A.G., Press W.H. \& Kirschner R.P., 1995,
 {\it ApJ} {\bf 445}, L91.\\
 Roulet E. \& Tommasini D., 1991, {\it Phys. Lett.}
{\bf B 256}, 218.\\
Savaglio S., Cristiani S., D'Odorico S., Fontana A., Giallongo E. \& Molaro
 P., 1996,\\ \indent ASTRO-PH 9606063, A\&A (in press).\\
 Sciama D.W., 1990a, {\it Phys. Rev.
Lett.} {\bf 65}, 2839.\\ Sciama D.W., 1990b, {\it ApJ} {\bf 364}, 549.\\
Sciama D.W., 1991, {\it ApJ} {\bf 367}, L39.\\ Sciama D.W., 1993, {\it
Modern Cosmology and the Dark Matter problem}\\ \indent (Cambridge
University Press).\\ Sciama D.W., 1994, {\it ApJ} { \bf 422}, L49.\\
Sciama D.W., 1995, {\it ApJ} {\bf 448}, 667.\\
 Sethi S.K., 1995, ASTRO-PH 9507116.\\
Shapiro P.R., 1995, {\it The physics of the
 interstellar medium and intergalactic medium}\\ \indent ASP Conference
Series {\bf 80}, 55.\\
Smette A., Robertson J.G., Shaver P.A., Reimers D., Wisotzki L. \& K\"ohler T.,
1995,\\ \indent A\&AS {\bf 113}, 199.\\
Songaila A., Hu E.M. \& Cowie L.L., 1995, {\it Nature} {\bf 375},124.\\
 Spitzer L., 1978, {\it Physical processes in the
interstellar medium}\\ \indent (John Wiley \& Sons, New York).\\
Steidel C., 1990, {\it ApJS} {\bf 74}, 37.
\\ Tammann G.A., Labhardt L., Federspiel M., Sandage A., Saha A., Macchetto F.D.
 \\
 \indent \& Panagia N., 1996, {\it Science with the Hubble Space Telescope},\\
 \indent Eds. P. Benvenuti, F.D. Macchetto \& J. Schreier.
\\ Tanvir N.R.,
Shanks T., Ferguson H.C. \& Robinson D.R.T., 1995, {\it Nature} {\bf 377}, 27.\\
Tytler D., Fan X.-M., Burles S., Cottrell L., David C., Kirkman D. \& Zuo
L.,\\ \indent 1995, {\it Proceedings of the ESO Workshop on Quasars
Absorption Lines},\\ \indent Ed. G. Meylan (Springer, Heidelberg) p. 289.\\
Williger G.M., Baldwin J.A., Carswell R.F., Cooke A.J., Hazard C., Irwin M.J.,\\
\indent McMahon R.G. \& Storrie-Lombardi L., 1994, {\it ApJ} {\bf 428}, 574.\\
Zanchin V., Lima J.A.S. \& Brandenberger R., 1996, ASTRO-PH 9607062.

\end{document}